\documentclass[twocolumn,prl,aps,amsfonts,amssymb,amsmath,showpacs]{revtex4}
\usepackage{graphicx}
\usepackage{hyperref}

\begin{document}

\title{``Negative'' Backaction Noise in Interferometric Detection of a Microlever}

\author{J. Laurent, A. Mosset, O. Arcizet, J. Chevrier, S. Huant, and H. Sellier}
\affiliation{Institut N\'eel, CNRS et Universit\'e Joseph Fourier, B.P. 166, F-38042 Grenoble Cedex
9, France}
\date{\today}

\begin{abstract}
Interferometric detection of mirror displacements is intrinsically limited by laser shot noise. In
practice, however, it is often limited by thermal noise. Here we report on an experiment performed
at the liquid helium temperature to overcome the thermal noise limitation and investigate the
effect of classical laser noise on a microlever that forms a Fabry-Perot cavity with an optical
fiber. The spectral noise densities show a region of ``negative'' contribution of the backaction
noise close to the resonance frequency. We interpret this noise reduction as a coherent coupling of
the microlever to the laser intensity noise. This optomechanical effect could be used to improve
the detection sensitivity as discussed in proposals going beyond the Standard Quantum Limit.
\end{abstract}

\pacs{07.60.Ly, 07.79.Lh, 42.50.Wk}
\maketitle


Coupling mechanical resonators with light has become an exciting field of research since the
discovery of the electromagnetic damping effect in a microwave cavity with a movable wall
\cite{braginsky-70-jetp,favero-09-npho}. Extended to optical cavities with flexible mirrors, this
effect has been shown to induce self-oscillations \cite{dorsel-83-prl} or self-cooling
\cite{arcizet-06-nat,gigan-06-nat} depending on the cavity detuning from the optical resonance. A
similar cooling effect known as cold damping can be obtained with an active feedback technique
\cite{mancini-98-prl,cohadon-99-prl}. Cavity cooling now appears as a promising route to cool down
macroscopic oscillators into their quantum state by using a high-finesse cavity in the resolved
sideband regime to extract phonons with photons \cite{schliesser-08-natp}.

Interferometric optical cavities are also a focus of research on quantum limited measurements with
applications in metrology for gravitational-wave detectors \cite{reitze-08-npho}. An historically
important limit, called the Standard Quantum Limit (SQL), results from the compromise between the
reflected shot noise of photons and the uncorrelated backaction noise induced by radiation pressure
on cavity mirrors \cite{braginsky-68-jetp}. Different detection schemes have been proposed to
overcome this limit, including correlation of noise quadratures in detuned cavities to produce an
effective cancellation of the backaction noise
\cite{caves-81-prd,braginsky-99-pla,buonanno-01-prd,khalili-01-os,arcizet-06-pra,tsang-10-prl}. Up
to now, only proofs of principle have been achieved experimentally
\cite{mowlowry-04-prl,caniard-07-prl,verlot-10-prl,marino-10-prl} because the quantum shot noise is
masked by the mirror thermal noise at room temperature. To reach the ultimate quantum limit of an
interferometric detection, it is therefore indispensable, first, to lower the temperature and,
then, to apply a quantum limited detection scheme.

In this Letter, we report advances in this direction by cooling down an optical cavity at 4.2\,K to
suppress the thermal noise. We use a microlever for Atomic Force Microscopy (AFM) as a flexible
mirror that forms a cavity with the extremity of an optical fiber \cite{rugar-89-apl}. Despite the
lower reflection coefficient as compared to the low-loss mirrors of quantum optics, microlevers
experience similar optomechanical effects due to retarded photothermal forces induced by light
absorption \cite{vogel-03-apl,metzger-04-nat}. In particular, self-cooling has been shown to reduce
the effective temperature, although it does not improve the signal-to-noise ratio for force
detection because the signal is damped in the same way as the thermal noise \cite{vitali-01-pra}.
Here, we demonstrate experimentally for the first time that a microlever, cooled at 4.2\,K to
suppress thermal noise, couples coherently to the classical intensity noise of the laser beam and
gives a reduction of the measurement noise. This ``negative'' contribution of the backaction noise
occurs just above or below the mechanical resonance frequency depending on cavity detuning. This
noise reduction effect, demonstrated here in the classical regime on a simple system, represents
one of the proposed schemes to beat the SQL when applied in the quantum regime of shot noise.


We perform the experiments in a cryogenic force microscope at 4.2\,K under a low pressure of helium
gas for thermalization. The microlever is a commercial silicon cantilever (230\,$\mu$m long,
40\,$\mu$m wide, 3\,$\mu$m thick) coated with 80\,nm of gold on the interferometer side and 200\,nm
on the tip side. The spring constant $K=8$\,N/m is determined from thermal noise spectra at 300\,K
when optomechanical effects are negligible \cite{hutter-93-rsi}. The fundamental resonance
frequency is 41555\,Hz and the quality factor reaches 15000 at 4.2\,K.

\begin{figure}
\begin{center}
\includegraphics[width=\columnwidth,clip,trim=0 28 13 0]{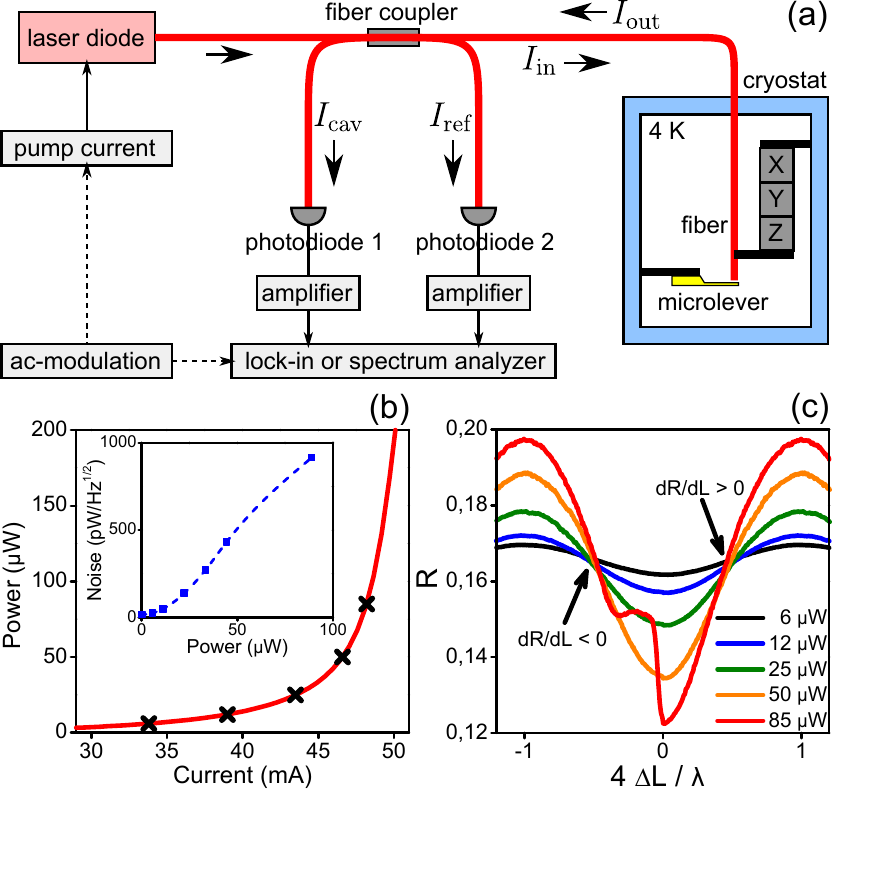}
\caption{(a) Experimental setup. (b) Light intensity versus pump current showing the lasing
threshold around 45\,mA. Markers indicate the powers used in Figs.~\ref{figure1}(c), \ref{figure2}
and \ref{figure3}. Inset: Laser noise versus laser power. (c) Reflection coefficient $R$ versus
cavity length detuning $\Delta L$ for several incident powers (labels). Detunings used for
interferometric detection are indicated by arrows.} \label{figure1}
\end{center}
\end{figure}

The light source is a $\lambda=670$\,nm laser diode stabilized in temperature and protected by a
Faraday isolator. The laser is coupled to a single-mode optical fiber connected to a 50\%-50\%
fiber coupler [Fig.~\ref{figure1}(a)]. One of the output fibers is sent into the cryostat to
measure the microlever motion by interferometry in the parallel cavity formed by the lever and the
cleaved end of the fiber, coated with 15\,nm of gold to increase the reflection coefficient
(16\,\%). The remaining ports of the coupler are connected to two photodiodes (0.44\,A/W) with
low-noise current amplifiers, one recording the same intensity as the incident light
($I_{\text{ref}}=I_{\text{in}}$) and one recording half of the reflected intensity
($I_{\text{cav}}=I_{\text{out}}/2$).

The fiber is approached at a few tens of microns from the lever, and the light beam is centered on
the lever end by using inertial motors. The Z translation stage of the fiber is used to scan the
cavity length $L$ and record the interference pattern of the reflection coefficient
$R=I_{\text{out}}/I_{\text{in}}$, as shown in Fig.~\ref{figure1}(c) for five powers obtained by
changing the pump current on the laser diode [Fig.~\ref{figure1}(b)]. The higher contrast at a
large power results from the increasing coherence length around the lasing threshold. The detection
of the lever motion is done at the middle of the interference pattern where the slope
$\frac{dR}{dL}$ is maximum, positive or negative. By fitting the interference patterns with the
Fabry-Perot formula, one gets a cavity finesse $\mathcal{F}$ ranging from 0.4 at 6\,$\mu$W to 1.2
at 85\,$\mu$W, suggesting that multiple reflections inside the cavity are limited. On the other
hand, the spontaneous oscillations observed at a large power (85\,$\mu$W) on the negative slope
(i.e., negative detuning), and visible as a kink on the bottom-left side of the interference
pattern, show that the intracavity field is strongly sensitive to the cavity length, a situation
that usually requires a good finesse \cite{favero-09-npho}. In addition, the instability does not
appear on the steepest part of the pattern but close to zero detuning. It seems therefore that the
reflection coefficient measured through the small fiber core does not represent the actual finesse
of the cavity, possibly due to the diverging light beam in the larger cavity formed by the fiber
cladding.

\begin{figure}
\begin{center}
\includegraphics[width=\columnwidth,clip,trim=13 13 20 20]{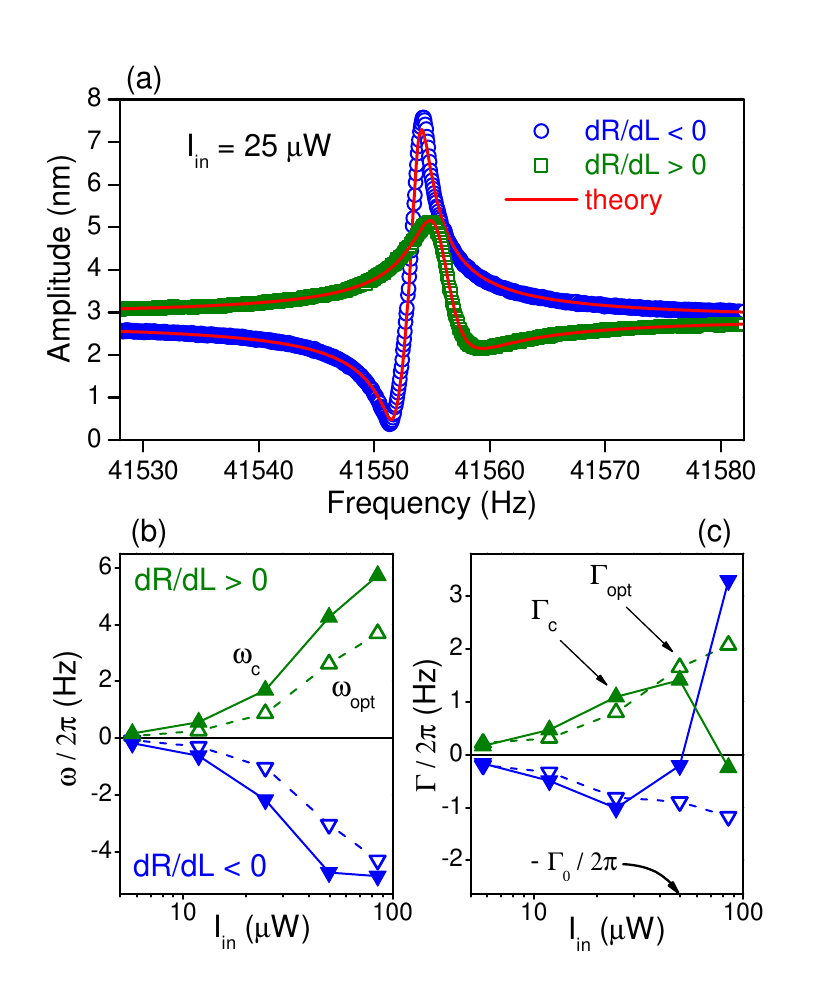}
\caption{(a) Microlever frequency response to light intensity modulation at 25\,$\mu$W incident
power for positive and negative cavity slopes. Solid lines are fitting curves with function
$|\eta(\omega)|$. (b),(c) Parameters obtained by fitting response curves at different powers.
$\omega_c/2\pi$ and $\Gamma_c/2\pi$ describe the shape of the effective reflection coefficient
$\eta(\omega)$ (closed symbols). $\omega_{\text{opt}}/2\pi$ and $\Gamma_{\text{opt}}/2\pi$ describe
the resonance frequency and damping rate changes (open symbols) around the intrinsic values
$\omega_0/2\pi=41554.77$\,Hz and $\Gamma_0/2\pi=2.64$\,Hz (position of the bottom axis) induced by
the optical spring effect.} \label{figure2}
\end{center}
\end{figure}

In order to gain physical insight into the origin of the optomechanical coupling, we first analyze
the lever response to a small sinusoidal modulation of the laser pump current ($\delta
i/i=4\times10^{-4}$) resulting in a modulation of the incident light intensity. The modulation of
the intensity reflected by the cavity is recorded as a function of the modulation frequency with a
lock-in amplifier, for a given average power and for a positive or negative detuning. The response
for an intermediate power of 25\,$\mu$W is plotted in Fig.~\ref{figure2}(a) and shows an asymmetric
Fano-like resonance because the flat response of the direct reflection is coupled to the resonant
response of the lever \cite{kadri-11-oe}. Since the phase of the lever vibration switches from 0 to
$-\pi$ across the resonance, the vibration signal is first added to and then subtracted from the
direct reflection, giving successively a peak and a dip in the total response, for a positive
slope. This relative position of the peak and dip indicates that the optical force pushes the lever
like a radiation pressure. For a negative slope, the sign of the vibration signal is reversed,
exchanging the peak and dip positions.

We model the optomechanical response around the resonance frequency as explained below. The light
intensity fluctuations $\delta I_{\text{in}}(\omega)$ at angular frequency $\omega$ induce
fluctuations in the optical force and result in fluctuations $\delta z(\omega)$ in the lever
position according to \cite{metzger-08-prb} :
$$\delta z(\omega)=\frac{\omega_0^2/K}{\omega_1^2-\omega^2+i\omega\Gamma_1}\;\frac{2\beta/c}{1+i\omega\tau}\;\delta I_{\text{in}}(\omega)$$
The first part is the lever response $\chi(\omega)$ involving the intrinsic spring constant $K$,
the resonant frequency $\omega_1=\omega_0+\omega_{\text{opt}}$ and the damping rate
$\Gamma_1=\Gamma_0+\Gamma_{\text{opt}}$. These parameters are modified with respect to the
intrinsic values $\omega_0$ and $\Gamma_0$ by the quantities
$\omega_{\text{opt}}=\frac{K_{\text{opt}}}{K}\frac{\omega_0/2}{1+\omega_0^2\tau_{\text{opt}}^2}$
and
$\Gamma_{\text{opt}}=-\frac{K_{\text{opt}}}{K}\frac{\omega_0^2\tau_{\text{opt}}}{1+\omega_0^2\tau_{\text{opt}}^2}$
related to the cavity-length-dependent optical force characterized by a spring constant
$K_{\text{opt}}$ and a force delay $\tau_{\text{opt}}$. The second part of the equation is the
optical force at angular frequency $\omega$ expressed in terms of a dimensionless parameter $\beta$
giving the strength of the actual optomechanical force relative to the ideal radiation pressure
induced by $\delta I_{\text{in}}(\omega)$. Note that $\beta$ is usually larger than 1 due to light
amplification in the cavity, but it can also represent a photothermal force instead of radiation
pressure. $\tau$ is the retardation of the force, and $c$ the speed of light.

The reflected light intensity $I_{\text{out}}=R(L)\;I_{\text{in}}$ has fluctuations $\delta
I_{\text{out}}=R\;\delta I_{\text{in}}+I_{\text{in}}\;\frac{dR}{dL}\;\delta z$ composed of
reflected fluctuations for fixed cavity length and cavity length fluctuations described by the
optomechanical response discussed above. These two contributions are added coherently and the
reflected fluctuations $\delta I_{\text{out}}(\omega)=R\;\eta(\omega)\;\delta
I_{\text{in}}(\omega)$ can be expressed in terms of an effective reflection coefficient for
intensity fluctuations characterized by a dimensionless function $\eta(\omega)$. For $\omega$
around the mechanical resonance, a large quality factor, and a small optical spring effect, this
function reduces to the simple form :
$$\eta(\omega)=\frac{(\omega_1+\omega_c-\omega)+i\;(\Gamma_1+\Gamma_c)/2}{(\omega_1-\omega)+i\;(\Gamma_1/2)}$$
where $\omega_c=C\omega_1/2$ and $\Gamma_c=-C\omega_1^2\tau$ can be expressed with the
dimensionless coefficient
$C=\frac{I_{\text{in}}}{R}\frac{dR}{dL}\frac{2\beta}{cK}\frac{1}{1+\omega_1^2\tau^2}$ describing
the coupling strength.

The experimental response curves are perfectly fitted with $|\eta(\omega)|$ as shown in
Fig.~\ref{figure2}(a). The fitting parameters are plotted for several powers and both cavity slopes
in Figs.~\ref{figure2}(b) and \ref{figure2}(c). The parameters $\omega_c$ and $\omega_{\text{opt}}$
have the same sign and share the same evolution with incident power, suggesting a single
optomechanical coupling for the response to the modulation and the optical spring effect. They are,
respectively, proportional to the derivative (with respect to the cavity length) of the reflected
and intracavity intensities, which have similar values here because of the low cavity finesse. The
positive sign obtained for a positive cavity slope would be consistent with a coupling by radiation
pressure, but the measured value of $C=2\omega_c/\omega_1\sim 10^{-4}$ at 25\,$\mu$W together with
$\tau\sim 0$ for radiation pressure gives $\beta\sim 2000$ corresponding to a very large
enhancement of the intracavity field which is not realistic for a low cavity finesse. As a result,
the optomechanical force is likely of photothermal origin, as expected for a lever coated with
asymmetric metallic layers.

At a low power, the signs of $\Gamma_c$ and $\Gamma_{\text{opt}}$ are the same as for $\omega_c$
and $\omega_{\text{opt}}$ on the same slope. Since theory predicts opposite signs for $\omega$ and
$\Gamma$ parameters, the coupling probably involves two optical forces with opposite directions
\cite{metzger-04-nat}, one with a large delay $\tau$ giving a larger $\Gamma_c$ and one with a
small delay $\tau$ giving a larger $\omega_c$. This situation is possible in the case of two
photothermal forces arising from the coexistence of two heat conduction paths with different time
scales inside the lever, producing mechanical stresses in opposite directions. The same experiment
performed at 300\,K finds a much weaker optomechanical coupling, due possibly to a different
thermomechanical response. Note that we cannot determine independently the sign and delay of the
two photothermal forces by recording the imaginary part of the response on a large frequency range
\cite{metzger-08-prb} because the response is below the detection sensitivity out of resonance.

Remarkably, $\Gamma_c$ reverses sign at a large power, suggesting that several photothermal forces
contribute with different dependencies on incident power, but the origin of all these forces is not
clearly understood. On the other hand, $\Gamma_{\text{opt}}$ keeps a constant sign and would
eventually reach $-\Gamma_0/2\pi$ at an even larger power, leading to the self-oscillations
observed for negative detuning. This difference between $\Gamma_c$ and $\Gamma_{\text{opt}}$ power
behaviors might result from the difference between the reflected and intracavity fields they are
connected to, in particular around the lasing threshold where the coherence length suddenly
increases.

\begin{figure}
\begin{center}
\includegraphics[width=\columnwidth,clip,trim=10 10 15 15]{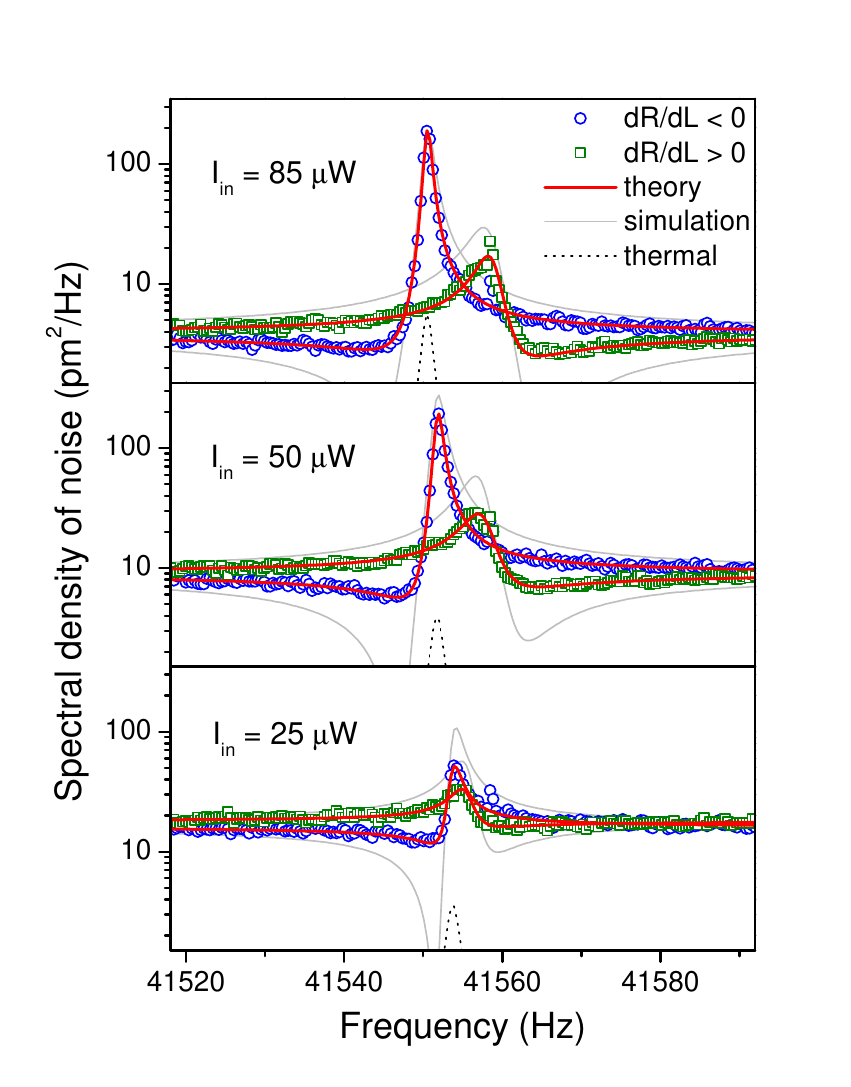}
\caption{Power spectral density of the reflected light intensity $S_{\text{out}}(\omega)$ measured
for a negative (blue circles) and positive (green squares) slope of the interference pattern and
three incident powers. Red solid lines are fitting curves using $|\eta(\omega)|^2$. Gray solid
lines are simulations using $|\eta(\omega)|^2$ with the parameters of Figs.~\ref{figure2}(b) and
\ref{figure2}(c). Gray dotted lines are thermal noise spectra expected at 4.2\,K (for a negative
slope, see the text).} \label{figure3}
\end{center}
\end{figure}

Having explored the optomechanical coupling in the presence of an external modulation, we now turn
to the analysis of the noise spectrum in the absence of excitation. The incident light intensity
emitted by the laser diode has time fluctuations characterized by a flat power spectral density
$S_{\text{in}}(\omega)$ from zero to above 50\,kHz. This white classical noise increases with power
as shown in the inset in Fig.~\ref{figure1}(b). The power spectral density $S_{\text{out}}(\omega)$
of the reflected light is converted into lever displacement noise and plotted in Fig.~\ref{figure3}
for three powers. All these spectra show an asymmetric shape and a local minimum, which have never
been observed, to the best of our knowledge, in noise spectra of microlevers, since they are
usually dominated by thermal noise. Here, the spectra contain the flat contribution due to the
\textit{direct} reflection of the incident noise (as for a rigid cavity) and the resonant
contribution due to the \textit{backaction} on the microlever induced by photothermal coupling
\cite{braginsky-99-pla2,cerdonio-01-prd}. These two noises dominate by a factor of 10 over the
\textit{detection} noise of the photodiode amplifier and over the \textit{thermal} noise at 4\,K.
Since the \textit{backaction} noise is correlated to the \textit{direct} noise, the spectrum is an
asymmetric resonance with a region of reduced noise as compared to the uncorrelated situation.
Moreover, the presence of a dip in the spectrum shows that the \textit{backaction} noise acts as a
``negative'' noise on top of the flat \textit{direct} noise. The frequency range of the reduced
noise is located on the left (right) side of the resonance for negative (positive) slope of the
cavity in agreement with the intensity modulation experiment. The peak height difference between
positive and negative slopes results from the retarded optical spring force in the cavity (damping
rate change $\Gamma_{\text{opt}}$) and is equivalent to the self-cooling and self-heating effects
observed in the case of thermal noise \cite{metzger-04-nat}.

According to the model, the noise in the reflected intensity is given by
$S_{\text{out}}(\omega)=R^2\;|\eta(\omega)|^2\;S_{\text{in}}(\omega)$. The experimental spectra are
correctly fitted with this expression, though with different fitting parameters from those of the
modulation experiment which give deeper minima (see Fig.~\ref{figure3}). The fitting parameters
$\omega_{\text{opt}}$ and $\Gamma_{\text{opt}}$ are almost the same in both experiments, but
$\omega_c$ is about 3 times lower in the noise spectra and $\Gamma_c$ is also quite different. This
difference could indicate that other sources of fluctuations might contribute to the spectra. The
absence of excess noise on the slopes of the interference pattern shows, however, that the laser
\textit{frequency} noise is negligible. The photodiode \textit{detection} noise also has a much
lower level. The microlever \textit{thermal} noise calculated at 4.2\,K with the effective
parameters $\omega_{\text{opt}}$ and $\Gamma_{\text{opt}}$ \cite{metzger-08-prb} is also negligible
and even not visible in Fig.~\ref{figure3} for the positive slope because of self-cooling.
Alternatively, a possible explanation could be that the pump current modulation, used to create the
intensity modulation in the first experiment, also produces an optical frequency modulation, which
is then converted by the cavity into a modulation of the reflected intensity but uncoupled from the
lever dynamics.

The key fact, however, is that the spectral density shows a minimum of noise lower than the direct
intensity noise of the interferometric detection. Contrary to intuition, the \textit{backaction}
noise can therefore improve the signal-to-noise ratio of a weak force measurement in dynamic mode
by choosing a frequency slightly off resonance. Since the spectral region of reduced noise is
limited to a few hertz and near a region of enhanced noise, the bandwidth of the lock-in detection
will be limited to these few hertz of reduced noise around the maximum of the signal-to-noise ratio
given by $|\chi(\omega)|/|\eta(\omega)|$. According to this formula, the lowest noise is obtained
at angular frequency $\omega\approx\omega_1+\omega_c$ and could be lowered down to zero for
$\Gamma_c=-\Gamma_1$ if the quantum mechanical uncertainty principle would not set a lower bound on
the measurement noise \cite{tsang-11-prl}. Note that large $\Gamma_c$ values are obtained for the
optimum delay $\tau$ given by $\omega_1\tau\sim 1$.

Recently, a few quantum optics experiments
\cite{mowlowry-04-prl,caniard-07-prl,verlot-10-prl,marino-10-prl} have discussed the possibility to
beat the SQL by using an artificial noise above the thermal noise at 300\,K and a coupling by
radiation pressure. Our work now demonstrates that the backaction noise can be used to compensate
coherently for the reflected intensity noise at a particular frequency with such a simple system as
an AFM cantilever. This optomechanical effect will be useful to improve the sensitivity of
lever-based force detection experiments. Demonstrated here on a microlever cooled at the liquid
helium temperature and coupled to classical laser noise by photothermal forces, we expect this
effect to show up similarly with radiation pressure and in the quantum limit of photon shot noise.

\acknowledgments

This research was supported by the ANR PNANO 2006 program under the project name ``MONACO''.




\begin{thebibliography}{99}

\bibitem{braginsky-70-jetp}
V.B.~Braginsky, A.B.~Manukin, and M.Yu.~Tikhonov, Sov. Phys. JETP {\bf 31}, 829 (1970).

\bibitem{favero-09-npho}
I.~Favero and K.~Karrai, Nat. Photon. {\bf 3}, 201 (2009).

\bibitem{dorsel-83-prl}
A.~Dorsel, {\it et al.}, Phys. Rev. Lett. {\bf 51}, 1550 (1983).

\bibitem{arcizet-06-nat}
O.~Arcizet, {\it et al.}, Nature {\bf 444}, 71 (2006).

\bibitem{gigan-06-nat}
S.~Gigan, {\it et al.}, Nature {\bf 444}, 67 (2006).

\bibitem{mancini-98-prl}
S.~Mancini, D.~Vitali, and P.~Tombesi, Phys. Rev. Lett. {\bf 80}, 688 (1998).

\bibitem{cohadon-99-prl}
P.~Cohadon, A.~Heidmann, and M.~Pinard, Phys. Rev. Lett. {\bf 83}, 3174 (1999).

\bibitem{schliesser-08-natp}
A.~Schliesser, {\it et al.}, Nat. Phys. {\bf 4}, 415 (2008).

\bibitem{reitze-08-npho}
D.~Reitze, Nat. Photon. {\bf 2}, 582 (2008).

\bibitem{braginsky-68-jetp}
V.B.~Braginsky, Sov. Phys. JETP {\bf 26}, 831 (1968).

\bibitem{caves-81-prd}
C.M.~Caves, Phys. Rev. D {\bf 23}, 1693 (1981).

\bibitem{braginsky-99-pla}
V.B.~Braginsky and F.Ya.~Khalili, Phys. Lett. A {\bf 257}, 241 (1999).

\bibitem{buonanno-01-prd}
A.~Buonanno and Y.~Chen, Phys. Rev. D {\bf 64}, 042006 (2001).

\bibitem{khalili-01-os}
F.Ya.~Khalili, Opt. Spectrosc. {\bf 91}, 542 (2001).

\bibitem{arcizet-06-pra}
O.~Arcizet, {\it et al.}, Phys. Rev. A {\bf 73}, 033819 (2006).

\bibitem{tsang-10-prl}
M.~Tsang and C.M.~Caves, Phys. Rev. Lett. {\bf 105}, 123601 (2010).

\bibitem{mowlowry-04-prl}
C.M.~Mow-Lowry, {\it et al.}, Phys. Rev. Lett. {\bf 92}, 161102 (2004).

\bibitem{caniard-07-prl}
T.~Caniard, {\it et al.}, Phys. Rev. Lett. {\bf 99}, 110801 (2007).

\bibitem{verlot-10-prl}
P.~Verlot, {\it et al.}, Phys. Rev. Lett. {\bf 104}, 133602 (2010).

\bibitem{marino-10-prl}
F.~Marino, {\it et al.}, Phys. Rev. Lett. {\bf 104}, 073601 (2010).

\bibitem{rugar-89-apl}
D.~Rugar, H.J.~Mamin, and P.~Guethner, Appl. Phys. Lett {\bf 55}, 2588 (1989).

\bibitem{vogel-03-apl}
M.~Vogel, {\it et al.}, Appl. Phys. Lett. {\bf 83}, 1337 (2003).

\bibitem{metzger-04-nat}
C.~Metzger and K.~Karrai, Nature {\bf 432}, 1002 (2004).

\bibitem{vitali-01-pra}
D.~Vitali, S.~Mancini, and P.~Tombesi, Phys. Rev. A {\bf 64}, 051401(R) (2001).

\bibitem{hutter-93-rsi}
J.L.~Hutter and J.~Bechhoefer, Rev. Sci. Instrum. {\bf 64}, 1868 (1993).

\bibitem{kadri-11-oe}
S.~Kadri, H.~Fujiwara, and K.~Sasaki, Opt. Express {\bf 19}, 2317 (2011).

\bibitem{metzger-08-prb}
C.~Metzger, {\it et al.}, Phys. Rev. B {\bf 78}, 035309 (2008).

\bibitem{braginsky-99-pla2}
V.B.~Braginsky, M.L.~Gorodetsky, and S.P.~Vyatchanin, Phys. Lett. A {\bf 264}, 1 (1999).

\bibitem{cerdonio-01-prd}
M.~Cerdonio, {\it et al.}, Phys. Rev. D {\bf 63}, 082003 (2001).

\bibitem{tsang-11-prl}
M.~Tsang, H.M.~Wiseman, and C.M.~Caves, Phys. Rev. Lett. {\bf 106}, 090401 (2011).


\end{thebibliography}
\end{document}